# Landau Levels of Bilayer Graphene in a WSe$_2$/Bilayer Graphene van der Waals Heterostructure

Ya-Wen Chuang[1], Jing Li[1], Hailong Fu[1], Kenji Watanabe[2], Takashi Taniguchi[2], and Jun Zhu[1,3]

[1]*Department of Physics, The Pennsylvania State University, University Park, Pennsylvania 16802, USA*
[2]*National Institute for Materials Science, 1-1 Namiki, Tsukuba 305-0044, Japan*
[3]*Center for 2-Dimensional and Layered Materials, The Pennsylvania State University, University Park, Pennsylvania 16802, USA*

**Abstract**
Heterostructures formed between two different van der Waals materials enable interactions and functionalities absent in each component. In this work we show that vicinity to an atomically thin WSe$_2$ sheet dramatically impacts the energies of the symmetry-broken low Landau levels of bilayer graphene, possibly due to Coulomb screening. We present a systematic study of the magnetic field and electrical displacement field dependences of the Landau level gaps at filling factor $\nu$ = 1, 2, 3, and compare to BN encapsulated pristine bilayer graphene. The exchange-dominated energy splitting between the $N$ = 0 and 1 orbital wave functions is significantly enhanced, which leads to a modified phase diagram at filling factor $\nu$ = 0 and larger energy gaps at $\nu$ = 1 and 3 in WSe$_2$/bilayer graphene heterostructures. The exchange-enhanced spin gap at $\nu$ = 2, on the other hand, is reduced by approximately two-fold. Our results demonstrate a possible way to engineer quantum Hall phenomena via van der Waals heterostructures.

## I. INTRODUCTION

The ability to form close coupling between dissimilar van der Waals materials opens the door to engineer new interactions and device functionalities absent in individual components [1-6]. Through proximity coupling, graphene can become superconducting [7], magnetic [8, 9] or acquire spin orbit coupling (SOC) strength to develop topological band structures or spintronic applications[10-17]. Engineering the dielectric environment provides an effective means to control Coulomb interactions and reduce charged impurity scattering [18, 19]. In atomically thin transition metal dichalcogenides, electron-electron interaction-dominated phenomena such as band gap and exciton binding energy are particularly sensitive to dielectric engineering. There, even the presence of a monolayer graphene capping layer has been shown to cause a significant reduction in the band gap [20, 21].

In a magnetic field, bilayer graphene (BLG) possesses eight closely spaced Landau levels near zero energy, which originate from the spin, valley, and orbital electronic degrees of freedom [22]. The $E$ = 0 octet (−4 < $\nu$ < 4) exhibits a complex Landau level (LL) structure and fascinating quantum Hall ferromagnetism as a result of competing many-body interactions [23-34]. A particularly intriguing and challenging aspect of the BLG LLs is $E_{10}$, the energy splitting between the $N$ = 0 and 1 orbital wave functions that are illustrated in Fig. 1(a). The $N$ = 0 orbitals, labeled as $|\pm 0\rangle$ states occupy the top/bottom layer solely while the $|\pm 1\rangle$ states have a small component occupying the opposite layer. Here, "+/−" signs denote states in K/K' valleys respectively. Each of the four states above has two spin configurations corresponding to $\sigma$ =↑ or ↓. Because of the different wave function distribution of the two orbitals, interlayer potential $U$ [35] and the interlayer Slonczewski-Weiss-McClure (SWM) hopping parameter $\gamma_4$ [36] contribute to $E_{10}$. Moreover, exchange corrections originating from the filled LLs strongly renormalize the energies of the two levels [24, 32, 37, 38]. The resulting $E_{10}$ is notoriously difficult to calculate but plays an important role in determining the nature of the ground states in the fractional quantum Hall regime [39-41]. Experiments show that $E_{10}$ is small and depends strongly on both the magnetic field $B$ and the perpendicular electric field $D$ but measurements have been incomplete [29, 30, 33, 42]. BLG samples used in these experiments are typically sandwiched between hexagonal Boron Nitride (BN) layers to obtain high quality. Heterostructures incorporating other van der Waals materials have not played a role in studying the LL physics of bilayer graphene.

We have fabricated dual-gated, BN encapsulated $WSe_2$/BLG heterostructures to explore proximity-induced SOC and the effect of dielectric screening on the LLs of BLG. Remarkably, we found that the presence of even a monolayer $WSe_2$ sheet has a significant impact on the LL energies of the BLG. The orbital energy splitting $E_{10}$, which manifests as the LL gap $\Delta_1$ and $\Delta_3$ at filling factors $\nu = 1$ and 3, is significantly enhanced compared to pristine BN encapsulated BLG. Meanwhile, the LL gap $\Delta_2$ at $\nu = 2$ is reduced by approximately two-fold. We performed a systematic study of the $D$- and $B$-dependences of $\Delta_{1,2,3}$ and a quantitative comparison to the corresponding energies in pristine BLG. Our results provide fresh experimental insights to understand the interaction-driven LLs in bilayer graphene and offer a potential pathway of tuning quantum Hall physics via van der Waals heterostructures. Extensive measurements on our devices have not uncovered significant proximity-induced SOC. Results are briefly discussed in Appendix B.

## II. SAMPLE PREPARATION

$WSe_2$/BLG devices are fabricated through a multistep process. BLG flakes are exfoliated from Kish graphite onto a 290 nm $SiO_2$/doped Si wafer coated with poly-propylene carbonate (PPC) and identified optically. The PPC film carrying the BLG flake is then peeled off and placed onto a poly dimethyl siloxane (PDMS) stamp. The standard dry transfer technique [43] is used to transfer the BLG flake to a hexagonal Boron Nitride (BN) flake already exfoliated onto a $SiO_2$/doped Si wafer. We perform standard e-beam lithography and reactive ion etching to pattern the BLG flake into a Hall bar [44, 45] followed by annealing in Ar/$H_2$ at 450°C for 3 hours to remove the resist residue. A BN/$WSe_2$ stack is then assembled and transferred onto the BLG/BN device to cover half of the Hall bar as shown in Fig. 1(d). The $WSe_2$ sheet is a monolayer in device #4 and bilayer in devices #2 and #3. Finally, we use e-beam lithography and physical vapor deposition to make Cr/Au top contacts to the exposed BLG terminals and the top gate, which covers the entire rectangular area of the Hall bar. The doped Si serves as the back gate.

## III. MEASUREMENTS

A schematic sideview of our devices, a color-enhanced optical micrograph of device #2 and an atomic force microscope (AFM) image of the central area of the device are shown in Figs. 1(b)-(d). The area probed in transport measurements (between electrodes 6 and 2) appears mostly flat and bubble-free in the AFM image, suggesting a good van der Waals coupling. Measurements are performed using standard low-frequency lock-in four-probe configurations on both $WSe_2$-covered (hereby denoted as $WSe_2$/BLG) and uncovered BLG (pristine BLG) regions at varying temperatures down to 20 mK and magnetic fields up to 18 T. $T$ = 1.6 K unless otherwise mentioned. We use the top and bottom gates to independently control the carrier density $n$ and the electric displacement field $D$ of the BLG [33, 34]. The WSe$_2$ sheet is not electrically contacted. Figure 2(a) plots the resistance vs top gate voltage of the WSe$_2$/BLG side in device #2 at fixed Si back gate voltages V$_{bg}$ as labeled in the plot. This measurement allows us to track the V$_{tg}$-V$_{bg}$ gating relation of the charge neutrality point (CNP) and identify the D =0 point as the global minimum of the CNP resistance peak. Figure 2 (b) plots such gating relation obtained on both the pristine and the WSe$_2$/BLG side of device #2 and their respective D=0 positions. That fact that both gating relations follow the same slope indicates negligible effect of the WSe$_2$ on the capacitance of the top gate. This is not surprising given the atomic thickness of the WSe$_2$ sheets used in our devices. The Fermi level of the system remained inside the band gap of WSe$_2$ in our experiments. The $D$ = 0 positions on both sides of the device display only a small shift in V$_{tg}$. This indicates a negligible amount of charge transfer between the WSe$_2$ and the BLG sheets.

Our WSe$_2$/BLG devices exhibit field effect mobility of 30,000−50,000 cm$^2$/Vs and well-developed Shubnikov-de Haas oscillations in a magnetic field $B$. Figure 3(a) compares two magnetoresistance traces $R_{xx}(B)$ obtained on the WSe$_2$/BLG side of device #2 (red trace) and the pristine BLG side of device #3 (blue trace). Both traces are tuned to the same electron density $n$ = $2.7\times10^{12}$/cm$^2$ and the same displacement field $D$ = 95 mV/nm. Both oscillations start at $B \sim$ 1.2 T and are approximately sinusoidal, indicating high, comparable sample quality. A detailed comparison of the $\nu$ = 2 gap energy obtained on these two devices is shown in Fig. 5(a).

The first prominent effect of the WSe$_2$/BLG heterostructure manifests in the ($B$, $D$) phase diagrams of the $\nu$ = 0 state, where several phases with different order parameters have been observed in pristine BLG [23, 26-28, 42, 46]. The blue dashed lines in Fig. 3(b) plot the phase diagram obtained in Li *et al.* [33] on pristine BN encapsulated devices, where we have labeled the low-$D$ canted-antiferromagnetic (CAF) and the high-$D$ layer polarized (LP) phases. Phase diagrams obtained by other groups on BN-encapsulated BLG are in excellent agreement with ours [28, 29, 32, 47]. At the phase boundaries labeled as $D^*_h$ and $D^*_l$ (blue dashed lines), the $\nu$ = 0 gap closes and $R_{xx}(D)$ exhibits a local minimum. In pristine BLG, an intermediate phase region between $D^*_h$ and $D^*_l$ (gray shaded area in the plot) starts to appear at $B$ = 12 T and grows with increasing $B$ and $D$. As the inset to Fig. 3(c) shows, this phase is expected to be partially polarized in the orbital index $N$, and its area is directly related to the magnitude of $E_{10}$ [33]. The main panel of Fig. 3(c) plots several $R(D)$ traces taken at $\nu$= 0 on the WSe$_2$/BLG side of device #2. Each trace corresponds to a fixed $B$-field. The splitting of $D^*_h$ and $D^*_l$ occurs at $B \sim$ 4−5 T, much smaller than the onset field in pristine BLG[28, 29, 32, 47]. The resulting phase diagram is plotted in Fig. 3(b) using red symbols. In this diagram, the intermediate phase occupies an area significantly larger than its counterpart in pristine BLG. Interestingly, the expansion of the intermediate phase comes at purely the expense of the CAF phase while $D^*_h$, the phase boundary to the LP state, are nearly identical in pristine and WSe$_2$/BLG.

To further examine the influence of the WSe$_2$ on the LLs of BLG, we measure directly the LL gap energy $\Delta_{1,2,3}$ at $\nu$ = 1, 2 and 3 using the temperature dependence of the magnetoresistance $R_{xx}(T)$. Figure 4(a) shows a set of $R_{xx}(n)$ sweeps that span $\nu$ = 1, 2, and 3 at selected temperatures ranging 3−20 K. Here $D$ = 100 mV/nm and $B$ = 8.9 T. Figure 4(b) shows the Arrhenius plots for the three filling factors and the corresponding fits to $R_{xx} \propto \exp(-\frac{\Delta}{2k_BT})$, where $\Delta_{1,2,3}$ = 1.6, 4.3 and 1.6 meV respectively from the fits. We chose the temperature range where the background of $R_{xx}$ remains constant (indicated by bundle points on either side of the integer fillings). The fits are well behaved and insensitive to the exact $V_{tg}$ choice of the analysis. Similar measurements are repeated at other magnetic and displacement fields.

## IV. ANALYSIS AND DISCUSSION

Figure 5(a) plots the $B$-dependence of $\Delta_{1,2,3}$ on the WSe$_2$/BLG side of device #2, together with $\Delta_2$ on the pristine side of device #3 (black stars) and $\Delta_2$ reported in the literature (shaded areas). Our $\Delta_2$ ($B$) measurements yield a slope of 0.6 meV/Tesla for WSe$_2$/BLG (purple dashed line) and 1.4 meV/Tesla for pristine BLG (black line). This comparison offers strong evidence that the gap of $\nu = 2$ is significantly reduced in WSe$_2$/BLG. Results of $\Delta_2$ ($B$) reported in the literature exhibit larger spread due to quality variations and the lack of control in the $D$-field. Nonetheless, they are consistent with the results of our pristine BLG in slope and magnitude and are all significantly larger than $\Delta_2$ of our WSe$_2$/BLG devices. In contrast to the behavior of $\Delta_2$, $\Delta_{1,3}$ in WSe$_2$/BLG (inset of Fig. 5(a)) becomes measurable at $B \sim 6$ T and are considerably *larger* than values reported for pristine BLG [29, 30, 42]. The increase of $\Delta_{1,3}$ is consistent with the large splitting of $D^*_h$ and $D^*_l$ shown in Fig. 3(c); both reflect the enhancement of $E_{10}$ in WSe$_2$/BLG samples. Measurements in device #4 show qualitatively consistent results but with smaller gap energies. The analysis and comparison are shown in Appendix A.

Figure 5(b) plots the $D$-dependence of $\Delta_{1,2,3}$ on the WSe$_2$/BLG side of device #2 at a fixed $B =$ 8.9 T. Measurements of this kind are lacking in pristine BLG. $\Delta_3$ (open and solid blue symbols) exhibits a rapid rise with increasing $D$ then saturates at approximately $D^* \sim 75$ mV/nm to a magnitude of 1.3 meV. In comparison, both $\Delta_2$ (black symbols) and $\Delta_1$ (magenta symbols) are non-monotonic in $D$, with $\Delta_2$ showing a maximum and $\Delta_1$ showing a significant dip in the vicinity of $D^*$. These complex trends can in fact be well captured by an effective LL energy diagram Li *et al.* developed for pristine BLG [33]. The inset to Fig. 5(b) illustrates the order and crossings of the LLs in the vicinity of $D^*_h$, $D^*$, and $D^*_l$. Several features of the model, i.e. the simultaneous maxing of $\Delta_2$ and vanishing of $\Delta_1$ at $D^*$, the equal magnitude of $\Delta_1$ and $\Delta_3$ at $D > D^*_h$, and the monotonic $D$-dependence of $\Delta_3 = E_{10}$ are validated by our measurements. We conclude that LLs in WSe$_2$/BLG follow the same basic structure as in pristine BLG.

We fit the measurements of $\Delta_{1,2,3}$ obtained here to the effective model described in Li *et al.* [33] to examine the quantitative difference of the two systems. The fitting results are plotted as orange, dark olive and blue solid lines in Fig. 5(b). The overall agreement with data is very good. The blue line plots a polynomial fit to $E_{10}(D) = \Delta_3(D)$, the salient feature of which is the saturation of $\Delta_3$ at $D > D^*$. The orange line plots $\Delta_2$ (meV) $= 0.09D$ (mV/nm) $- 0.05B$ (Tesla), which represents the gap of $\nu = 2$ in the regime of $D < D^*$. In this regime, $\Delta_2$ is a valley splitting as the inset shows and is primarily given by the size of the $D$-field induced band gap of BLG. The coefficient 0.09 is determined from independent band gap measurements at $B = 0$ in this device using methods similar to that described in the Supplementary material of Ref. [33]. The dark olive line plots $\Delta_2$ (meV) $= 0.67B$ (Tesla) $- E_{10}(D)$, which fits the gap of $\nu = 2$ in the regime of $D > D^*$. In Fig. 5(b), $B = 8.9$ T. Using the same expression of $\Delta_2$ and setting $D = 100$ mV/nm, we obtain $\Delta_2(B) = 0.67B - 1.3$. This equation is plotted as a dark olive line in Fig. 5(a) and also provides an excellent description of the measurements there. Consistency checks such as this show the effective model captures data really well. As the inset to Fig. 5(b) shows, $\Delta_2$ in the regime of $D > D^*$ is mostly an exchange-enhanced Zeeman gap, with the size of the spin splitting given by the linear term $\Delta_s = 0.67B$. Here the coefficient 0.67 is significantly smaller than $\Delta_s = 1.7B$ obtained for pristine BLG in Ref. [33], indicating a much weaker exchange effect at $\nu = 2$ in WSe$_2$/BLG. This is opposite to the situation of $E_{10}$, which is significantly enhanced in WSe$_2$/BLG.

It is remarkable that close vicinity to an atomically thin sheet of WSe$_2$ can lead to such large and contrasting changes of the LL gaps in BLG. It is possible that the dielectric screening of the WSe$_2$ sheet plays a role. However, unlike a thick BN dielectric layer used in existing BN encapsulated devices, here the WSe$_2$ sheet is less than 1 nm in thickness so it primarily impacts the short-range components of the Coulomb interaction. Indeed, the physics of the $E = 0$ octet in BLG is well known to be sensitive to interactions at the lattice scale [26, 27], where the presence of the WSe$_2$ sheet may well have an impact. The modified phase diagram of $\nu = 0$ we obtained in Fig. 3(b) supports this hypothesis. At finite fillings, both the enhanced Zeeman gap at $\nu = 2$ and the orbital splitting $E_{10}$ arise from exchange effects. How they are modified in the presence of a thin WSe$_2$ layer requires more in-depth calculations to understand. In addition to screening, we have considered the effect of twist alignment between the two lattices. Because the WSe$_2$ lattice is 34% larger than that of graphene, the Moiré pattern formed between the two has a wavelength of less than 1 nm for any given twist angle [48]. Guided by optical images of the devices, we produced various alignment scenarios between the selenium sublattice of WSe$_2$ and the carbon sublattice of BLG. All show the majority of the Se and C atoms do not directly overlap. Thus, direct on-site interactions could not have been significant in our devices. Further, the magnitude of the interlayer hopping term $\gamma_4$ is important to the size of $E_{10}$. Whether $\gamma_4$ is altered by the presence of the WSe$_2$ sheet requires calculations to clarify. We hope that the comprehensive and high-quality data we obtained on the LL energy gaps motivate future calculations and experiments to answer these open questions and by doing so shed more light on the complex many-body effects in BLG and van der Waals heterostructures in general.

## V. CONCLUSION

In conclusion, we have performed a systematic study of the LL energy gaps at filling factors $\nu = 1, 2$ and 3 in WSe$_2$/bilayer graphene heterostructures. The presence of an atomically thin WSe$_2$ sheet gives rise to a significantly enhanced $N = 0$ and 1 orbital splitting $E_{10}$ and a significantly reduced exchange Zeeman gap at $\nu = 2$. These measurements provide fresh input to understand interaction-driven quantum Hall phenomena in bilayer graphene and point to the use of van der Waals heterostructures as an effective knob to tune the strength of the interactions.

## ACKNOWLEDGEMENT

Work at Penn State is supported by the NSF through NSF-DMR-1708972 and NSF-DMR-1506212. H. F. acknowledges the support of the Eberly Postdoc Fellowship. K.W. and T.T. acknowledge support from the Elemental Strategy Initiative conducted by the MEXT, Japan and the CREST (JPMJCR15F3), JST. Part of this work was performed at the NHMFL, which was supported by the NSF through NSF-DMR-1157490, NSF-DMR-1644779 and the State of Florida. We thank Herbert Fertig, Efrat Shimshoni, Ganpathy Murthy, Jaroslav Fabian, Martin Gmitra and Jing Shi for helpful discussions. We are grateful to J. I. A. Li and Yihang Zeng for sharing with us their experiences of dry transfer and to Hongwoo Baek of the NHMFL for experimental assistance.

## APPENDIX A: MEASUREMENTS on DEVICE#4

Figure 6(a) shows an AFM image of the central area of device #4. The BLG is partially covered by a monolayer WSe$_2$ sheet outlined in orange and we performed measurements on the WSe$_2$/BLG side of the device. Compared to device #2, more bubbles formed in the transfer process. The gating characteristics of the device is shown in Fig. 6(b), in styles similar to Fig. 2. Figures 7(a)-(c) compare the measured LL gap energies $\Delta_{1,2,3}$ in both devices as a function of $D$- and $B$-fields. In all three figures, device #4 displays trends similar to device #2 but with consistently smaller gap values. Since the gaps at $v$ = 1, 2, 3 are all sensitive to exchange effects, larger density inhomogeneity generated by the bubbles in #4 may have caused a reduction in the gap energies. Despite a stronger disorder, Fig. 7(d) shows that in device #4 signs of splitting between $D^*_h$ and $D^*_l$ of the $v$ = 0 state appear at $B$ = 6 T whereas even in very high quality pristine BLG samples, the splitting does not occur until $B \sim$ 12 T[32, 33]. This observation supports the enhancement of $E_{10}$ in device #4 also, which likely comes from modified Coulomb interactions.

## APPENDIX B: PROXIMITY-INDUCED SPIN ORBIT COUPLING

An original goal of our study is to examine the experimental signatures of spin orbit coupling (SOC) in BLG, introduced through proximity coupling to a transition metal dichalcogenide (TMD) material with large SOC strength such as WSe$_2$ or WS$_2$ [14-16]. Previous experiments have indeed identified the signatures of proximity-induced SOC in the splitting of the bands– manifesting as a beating pattern in the Shubnikov-de Haas (SdH) oscillations– and weak antilocalization [14, 15]. The magnitude of the induced SOC is found to be of order 10 meV. Guided by band structure calculations [16] and previous experiments [14, 15], We have carefully looked for these signatures in our devices over a wide range of carrier type and density and at varying $D$-fields, with parallel studies performed on the pristine BLG region of the devices. We have not observed evidence that points to significant proximity-induced SOC. At very low magnetic field, our devices exhibit weak localization signals consistent with BLG behavior in our devices and reported previously [49]. No weak antilocalization was found. No clear beating pattern was present in the SdH oscillations of our devices. Figure 8(a) shows a few examples in device #2 at hole carrier density $n_h$ = 3.5 − 4 × 10$^{12}$/cm$^2$ and selected $D$-fields from −200 mV/nm to +200 mV/nm. For comparison, we also show the illustrated band structure evolution calculated by Gmitra *et al.* [16]. The SdH oscillations in all four panels are similar to one another, without clear signs of beating. These observations point to a very small, if any, induced SOC in our devices. We suspect that the discrepancy arises from the different sample preparation processes used. Devices used by Wang *et al.* [14, 15] have the stacking order of graphene/TMD/SiO$_2$. The stack was "ironed" with an AFM tip scanning in contact mode to remove contaminations [15]. This process could have also pushed graphene closer to the TMD sheet. In our fabrication process, the van der Waals coupling of WSe$_2$ to the BLG is made at the last transfer step, before which the WSe$_2$ and the BLG flakes are coupled to two separate BN sheets. In a recent work, Yang *et al.* [17] showed that the immediate vicinity to a BN sheet draws the graphene slightly away from the TMD and this distance increase, though only a small fraction of an Angstrom, can have a dramatic effect on the proximity-induced SOC strength. The absence of an observable SOC here is perhaps related to a larger WSe$_2$-BLG distance in our devices caused by the fabrication process we used. This is a hypothesis that needs to be further examined.

**Figure captions**

FIG. 1. (a) Illustration of the wave function $|\xi N\sigma\rangle$ in the $E = 0$ octet of bilayer graphene. $\xi = +/-$, $N = 0/1$ and $\sigma = \uparrow$ or $\downarrow$ denote the valley (K/K'), orbital and spin indices of the wave function respectively. The interlayer hopping parameter $\gamma_4$ connects a dimered site with a non-dimered site. (b) Schematic sideview of our BN encapsulated $WSe_2$/BLG devices. (c) Color-enhanced optical micrograph of device #2. The BLG sheet is shaded in purple. The bilayer $WSe_2$ flake is outlined in black. The orange dashed and gray dash dotted lines trace out the edges of the top and bottom BN sheets respectively. (d) Atomic force microscope image of device #2. The $WSe_2$ sheet is outlined in black dotted lines. $R_{xx}$ measurements use electrodes 5 and 6 on $WSe_2$/BLG and electrodes 2 and 3 on pristine BLG.

FIG. 2. (a) Resistance vs top gate voltage $R_{65}(V_{tg})$ at fixed Si back gate voltage $V_{bg}$ as labeled in the plot. Data show traces taken on the $WSe_2$/BLG side of device #2. The pristine BLG side looks similar. The black dashed line tracks the CNPs. (b) The $V_{tg}$-$V_{bg}$ relation of the CNP on both the $WSe_2$/BLG and the pristine BLG sides of device #2 with the $D = 0$ positions marked in the plot. The black dashed line is a guide to the eye.

FIG. 3. (a) Magnetoresistance $R_{xx}(B)$ obtained on the pristine BLG side of device #3 (blue trace, left axis) and the $WSe_2$/BLG side of device #2 (red trace, right axis). Both devices are set to electron density $n = 2.7\times10^{12}$/cm$^2$ and displacement field $D = 95$ mV/nm. (b) The $(D, B)$ phase diagram of the $v = 0$ state in BN encapsulated pristine BLG (blue dashed lines, from Li *et al.* [33]), and in $WSe_2$/BLG (red symbols). (c) Semi-log $R_{xx}(D)$ of $v = 0$ at selected $B$-fields from 3 to 6 T as labeled in the plot. $D^*_h$ and $D^*_l$ denote the upper and lower coincidence $D$-fields respectively. Inset: Schematic diagram of the LLs near the coincidence points. The black dotted line represents the Fermi level of $v = 0$. From device #2. $T = 20$ mK.

FIG. 4. (a) $R_{xx}(V_{tg})$ at selected temperatures as labeled in the plot. $V_{bg}$ was swept simultaneously to follow a line of constant $D = 100$ mV/nm. $B = 8.9$ T. The arrows point to filling factors $v = 1, 2$, and 3 respectively. (b) Arrhenius plot of $R_{xx}(T)$ at $v = 1$ (magenta), 2 (black), and 3 (blue). Fits to the data (solid lines) yield gap energies $\Delta_1 = 1.6$ meV, $\Delta_2 = 4.3$ meV and $\Delta_3 = 1.6$ meV respectively. From device #2.

FIG. 5. (a) The magnetic field dependence of $\Delta_{1,2,3}$ at $v = 1$ (magenta triangles), 2 (black squares), and 3 (blue circles) on $WSe_2$/BLG of device #2 and $\Delta_2$ at $v = 2$ (black stars) on pristine BLG of device #3. The purple dashed lines have the slope of 0.6 meV/Tesla. The black solid line 1.4

meV/Tesla. The dark olive line plots $\Delta_2(B) = 0.67B - 1.32$. Inset: a magnified view of $\Delta_{1,3}$. $D = 100$ mV/nm for all measurements. Shaded areas represent $\Delta_2$ reported in the literature (grey for [30], light green for [42], peach for [29]). (b) The $D$-field dependence of $\Delta_{1,2,3}$. $B = 8.9$ T. Symbols follow (a). Solid/open data points are obtained from density/$D$-field sweeps respectively. The solid orange line plots $\Delta_v$ (meV) $= 0.09D$ (mV/nm) $- 0.05B$ (Tesla). The solid dark olive line plots $0.67B - E_{10}(D)$, where $B = 8.9$ T and $E_{10}(D)$ is given by the blue solid line with an empirical function of $E_{10}(D) = 4.1\times10^{-2}D - 4.9\times10^{-4}D^2 + 2.7\times10^{-6}D^3 - 5.6\times10^{-9}D^4$ for $D < 100$ mV/nm and $E_{10} = 1.3$ meV for $D > 100$ mV/nm. The magenta dashed line is a guide to the eye for the $v = 1$ gap. Inset: LLs near the coincidence points. The $v = 1$ gap (shaded in pink) closes at $D^*$.

FIG. 6. (a) AFM image of the central area of device #4, where the BLG is partially covered by a monolayer WSe$_2$ sheet outlined in orange. (b) Resistance vs top gate voltage $R_{56}(V_{tg})$ at fixed Si back gate voltage $V_{bg}$, from 4V (leftmost, lime) to -11V (rightmost, violet) in 1V steps. The black dashed line tracks the CNP of each curve and the D=0 position. Inset: The $V_{tg}$-$V_{bg}$ relation of the CNP.

FIG. 7. The Landau level gap energies $\Delta_{1,2,3}$ measured in devices #2 and 4 as a function of the $D$-field (a) and $B$-fields ((b) and (c)) as labeled. (d) compares log $R_{xx}(D)$ at $v = 0$ in devices #4 (solid traces, left axis) and #2 (dashed traces, right axis). In device #4, prominent shoulders that suggest the appearance of $D^*_l$ are observed near places where clear splittings of $D^*_h$ and $D^*_l$ are seen in device #2.

FIG. 8. (a) Magnetoresistance oscillations at hole density $3.5 - 4 \times 10^{12}$/cm$^2$ and $D = -200$ mV/nm, $-65$ mV/nm, $+65$ mV/nm, and $+200$ mV/nm as labeled in the plots. A slowly varying background has been removed from each trace. The positive direction of the $D$-field points from the WSe$_2$ to the BLG to follow the definition used by Gmitra *et al.* [16]. We calculate $D$ following the convention of the field. (b) illustrates the expected evolution of the band structure. Beating is expected for the bottom two panels band but not for the top two panels because the Fermi level $E_F$ is in the valence band.

**References:**

[1]     A. K. Geim, and I. V. Grigorieva, Van der Waals heterostructures, Nature **499**, 419 (2013).
[2]     C. H. Lee, G. H. Lee, A. M. van der Zande, W. C. Chen, Y. L. Li, M. Y. Han, X. Cui, G. Arefe, C. Nuckolls, T. F. Heinz, J. Guo, J. Hone, and P. Kim, Atomically thin p-n junctions with van der Waals heterointerfaces, Nature Nanotechnology **9**, 676 (2014).
[3]     F. H. L. Koppens, T. Mueller, P. Avouris, A. C. Ferrari, M. S. Vitiello, and M. Polini, Photodetectors based on graphene, other two-dimensional materials and hybrid systems, Nature Nanotechnology **9**, 780 (2014).
[4]     P. Rivera, J. R. Schaibley, A. M. Jones, J. S. Ross, S. F. Wu, G. Aivazian, P. Klement, K. Seyler, G. Clark, N. J. Ghimire, J. Q. Yan, D. G. Mandrus, W. Yao, and X. D. Xu, Observation of long-lived interlayer excitons in monolayer MoSe2-WSe2 heterostructures, Nature Communications **6**, 6242 (2015).
[5]     T. Roy, M. Tosun, X. Cao, H. Fang, D. H. Lien, P. D. Zhao, Y. Z. Chen, Y. L. Chueh, J. Guo, and A. Javey, Dual-Gated MoS2/WSe2 van der Waals Tunnel Diodes and Transistors, Acs Nano **9**, 2071 (2015).
[6]     Y. K. Luo, J. S. Xu, T. C. Zhu, G. Z. Wu, E. J. McCormick, W. B. Zhan, M. R. Neupane, and R. K. Kawakami, Opto-Valleytronic Spin Injection in Monolayer MoS2/Few-Layer Graphene Hybrid Spin Valves, Nano Lett. **17**, 3877 (2017).
[7]     G. H. Lee, and H. J. Lee, Proximity coupling in superconductor-graphene heterostructures, Rep Prog Phys **81**, 056502 (2018).
[8]     Z. Y. Wang, C. Tang, R. Sachs, Y. Barlas, and J. Shi, Proximity-Induced Ferromagnetism in Graphene Revealed by the Anomalous Hall Effect, Physical Review Letters **114**, 016603 (2015).
[9]     P. Wei, S. Lee, F. Lemaitre, L. Pinel, D. Cutaia, W. Cha, F. Katmis, Y. Zhu, D. Heiman, J. Hone, J. S. Moodera, and C. T. Chen, Strong interfacial exchange field in the graphene/EuS heterostructure, Nat Mater **15**, 711 (2016).
[10]    C. L. Kane, and E. J. Mele, Quantum spin Hall effect in graphene, Physical Review Letters **95**, 226801 (2005).
[11]    Z. H. Qiao, W. Ren, H. Chen, L. Bellaiche, Z. Y. Zhang, A. H. MacDonald, and Q. Niu, Quantum Anomalous Hall Effect in Graphene Proximity Coupled to an Antiferromagnetic Insulator, Physical Review Letters **112**, 116404 (2014).
[12]    W. Han, R. K. Kawakami, M. Gmitra, and J. Fabian, Graphene spintronics, Nature Nanotechnology **9**, 794 (2014).
[13]    M. Gmitra, and J. Fabian, Graphene on transition-metal dichalcogenides: A platform for proximity spin-orbit physics and optospintronics, Physical Review B **92**, 155403 (2015).
[14]    Z. Wang, D. K. Ki, H. Chen, H. Berger, A. H. MacDonald, and A. F. Morpurgo, Strong interface-induced spin-orbit interaction in graphene on WS2, Nature Communications **6**, 8339 (2015).
[15]    Z. Wang, D.-K. Ki, J. Y. Khoo, D. Mauro, H. Berger, L. S. Levitov, and A. F. Morpurgo, Origin and Magnitude of `Designer' Spin-Orbit Interaction in Graphene on Semiconducting Transition Metal Dichalcogenides, Physical Review X **6**, 041020 (2016).
[16]    M. Gmitra, and J. Fabian, Proximity Effects in Bilayer Graphene on Monolayer WSe2: Field-Effect Spin Valley Locking, Spin-Orbit Valve, and Spin Transistor, Physical Review Letters **119**, 146401 (2017).
[17]    B. W. Yang, E. Molina, J. Kim, D. Barroso, M. Lohmann, Y. W. Liu, Y. D. Xu, R. Q. Wu, L. Bartels, K. Watanabe, T. Taniguchi, and J. Shi, Effect of Distance on Photoluminescence Quenching and Proximity-Induced Spin-Orbit Coupling in Graphene/WSe$_2$ Heterostructures, Nano Lett. **18**, 3580 (2018).
[18]    M. Kumagai, and T. Takagahara, Excitonic and nonlinear-optical properties of dielectric quantum-well structures, Physical Review B **40**, 12359 (1989).
[19]    A. K. M. Newaz, Y. S. Puzyrev, B. Wang, S. T. Pantelides, and K. I. Bolotin, Probing charge scattering mechanisms in suspended graphene by varying its dielectric environment, Nature Communications **3**, 734 (2012).
[20]    Y. X. Lin, X. Ling, L. L. Yu, S. X. Huang, A. L. Hsu, Y. H. Lee, J. Kong, M. S. Dresselhaus, and T. Palacios, Dielectric Screening of Excitons and Trions in Single-Layer MoS2, Nano Lett. **14**, 5569 (2014).
[21]    A. Raja, A. Chaves, J. Yu, G. Arefe, H. M. Hill, A. F. Rigosi, T. C. Berkelbach, P. Nagler, C. Schüller, T. Korn, C. Nuckolls, J. Hone, L. E. Brus, T. F. Heinz, D. R. Reichman, and A. Chernikov, Coulomb engineering of the bandgap and excitons in two-dimensional materials, Nature Communications **8**, 15251 (2017).
[22]    E. McCann, and M. Koshino, The electronic properties of bilayer graphene, Reports on Progress in Physics **76**, 056503 (2013).
[23]    R. T. Weitz, M. T. Allen, B. E. Feldman, J. Martin, and A. Yacoby, Broken-Symmetry States in Doubly Gated Suspended Bilayer Graphene, Science **330**, 812 (2010).

[24] J. Jung, F. Zhang, and A. H. MacDonald, Lattice theory of pseudospin ferromagnetism in bilayer graphene: Competing interaction-induced quantum Hall states, Physical Review B **83**, 115408 (2011).
[25] J. Velasco Jr, L. Jing, W. Bao, Y. Lee, P. Kratz, V. Aji, M. Bockrath, C. N. Lau, C. Varma, R. Stillwell, D. Smirnov, F. Zhang, J. Jung, and A. H. MacDonald, Transport spectroscopy of symmetry-broken insulating states in bilayer graphene, Nature Nanotechnology **7**, 156 (2012).
[26] M. Kharitonov, Antiferromagnetic state in bilayer graphene, Physical Review B **86**, 195435 (2012).
[27] M. Kharitonov, Canted Antiferromagnetic Phase of the v=0 Quantum Hall State in Bilayer Graphene, Physical Review Letters **109**, 046803 (2012).
[28] P. Maher, C. R. Dean, A. F. Young, T. Taniguchi, K. Watanabe, K. L. Shepard, J. Hone, and P. Kim, Evidence for a spin phase transition at charge neutrality in bilayer graphene, Nature Physics **9**, 154 (2013).
[29] K. Lee, B. Fallahazad, J. Xue, D. C. Dillen, K. Kim, T. Taniguchi, K. Watanabe, and E. Tutuc, Chemical potential and quantum Hall ferromagnetism in bilayer graphene, Science **345**, 58 (2014).
[30] A. Kou, B. E. Feldman, A. J. Levin, B. I. Halperin, K. Watanabe, T. Taniguchi, and A. Yacoby, Electron-hole asymmetric integer and fractional quantum Hall effect in bilayer graphene, Science **345**, 55 (2014).
[31] J. Velasco, Y. Lee, F. Zhang, K. Myhro, D. Tran, M. Deo, D. Smirnov, A. H. MacDonald, and C. N. Lau, Competing ordered states with filling factor two in bilayer graphene, Nature Communications **5**, 4550 (2014).
[32] B. M. Hunt, J. I. A. Li, A. A. Zibrov, L. Wang, T. Taniguchi, K. Watanabe, J. Hone, C. R. Dean, M. Zaletel, R. C. Ashoori, and A. F. Young, Direct measurement of discrete valley and orbital quantum numbers in bilayer graphene, Nature Communications **8**, 948 (2017).
[33] J. Li, Y. Tupikov, K. Watanabe, T. Taniguchi, and J. Zhu, Effective Landau Level Diagram of Bilayer Graphene, Physical Review Letters **120**, 047701 (2018).
[34] J. Li, H. Fu, Z. Yin, K. Watanabe, T. Taniguchi, and J. Zhu, Metallic Phase and Temperature Dependence of the v=0 Quantum Hall State in Bilayer Graphene, Phys. Rev. Lett. **122**, 097701 (2019).
[35] E. McCann, and V. I. Fal'ko, Landau-level degeneracy and quantum hall effect in a graphite bilayer, Physical Review Letters **96**, 086805 (2006).
[36] J. Lambert, and R. Côté, Quantum Hall ferromagnetic phases in the Landau level N=0 of a graphene bilayer, Physical review B **87**, 115415 (2013).
[37] G. Murthy, E. Shimshoni, and H. A. Fertig, Spin-valley coherent phases of the v = 0 quantum Hall state in bilayer graphene, Physical Review B **96**, 245125 (2017).
[38] K. Shizuya, Structure and the Lamb-shift-like quantum splitting of the pseudo-zero-mode Landau levels in bilayer graphene, Physical Review B **86**, 045431 (2012).
[39] E. H. Rezayi, and S. H. Simon, Breaking of Particle-Hole Symmetry by Landau Level Mixing in the v=5/2 Quantized Hall State, Physical Review Letters **106**, 116801 (2011).
[40] Z. Papic, and D. A. Abanin, Topological Phases in the Zeroth Landau Level of Bilayer Graphene, Physical Review Letters **112** (2014).
[41] A. A. Zibrov, C. Kometter, H. Zhou, E. M. Spanton, T. Taniguchi, K. Watanabe, M. P. Zaletel, and A. F. Young, Tunable interacting composite fermion phases in a half-filled bilayer-graphene Landau level, Nature **549**, 360 (2017).
[42] J. Martin, B. E. Feldman, R. T. Weitz, M. T. Allen, and A. Yacoby, Local compressibility measurements of correlated states in suspended bilayer graphene, Physical review letters **105**, 256806 (2010).
[43] L. Wang, I. Meric, P. Y. Huang, Q. Gao, Y. Gao, H. Tran, T. Taniguchi, K. Watanabe, L. M. Campos, D. A. Muller, J. Guo, P. Kim, J. Hone, K. L. Shepard, and C. R. Dean, One-dimensional electrical contact to a two-dimensional material, Science **342**, 614 (2013).
[44] J. Li, H. Wen, K. Watanabe, T. Taniguchi, and J. Zhu, Gate-Controlled Transmission of Quantum Hall Edge States in Bilayer Graphene, Physical Review Letters **120**, 057701 (2018).
[45] J. Li, K. Wang, K. J. McFaul, Z. Zern, Y. Ren, K. Watanabe, T. Taniguchi, Z. Qiao, and J. Zhu, Gate-controlled topological conducting channels in bilayer graphene, Nature Nanotechnology **11**, 1060 (2016).
[46] W. Bao, J. Velasco, F. Zhang, L. Jing, B. Standley, D. Smirnov, M. Bockrath, A. H. MacDonald, and C. N. Lau, Evidence for a spontaneous gapped state in ultraclean bilayer graphene, Proceedings of the National Academy of Sciences **109**, 10802 (2012).
[47] P. Maher, L. Wang, Y. D. Gao, C. Forsythe, T. Taniguchi, K. Watanabe, D. Abanin, Z. Papic, P. Cadden-Zimansky, J. Hone, P. Kim, and C. R. Dean, Tunable fractional quantum Hall phases in bilayer graphene, Science **345**, 61 (2014).
[48] S. Tang, H. Wang, Y. Zhang, A. Li, H. Xie, X. Liu, L. Liu, T. Li, F. Huang, X. Xie, and M. Jiang, Precisely aligned graphene grown on hexagonal boron nitride by catalyst free chemical vapor deposition, Scientific Reports **3**, 2666 (2013).

[49]     R. V. Gorbachev, F. V. Tikhonenko, A. S. Mayorov, D. W. Horsell, and A. K. Savchenko, Weak localization in bilayer graphene, Physical Review Letters **98**, 176805 (2007).

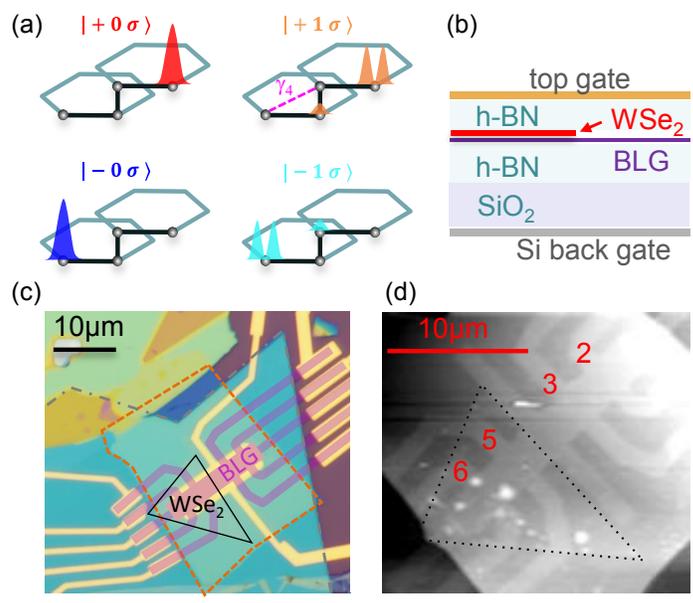

Figure 1

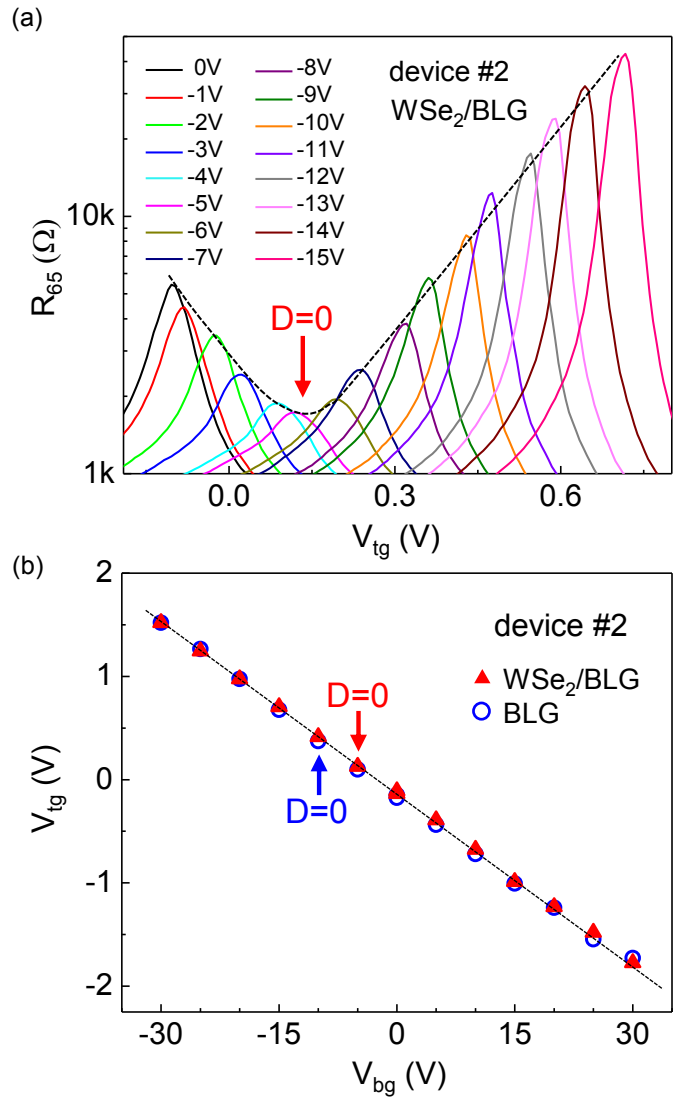

Figure 2

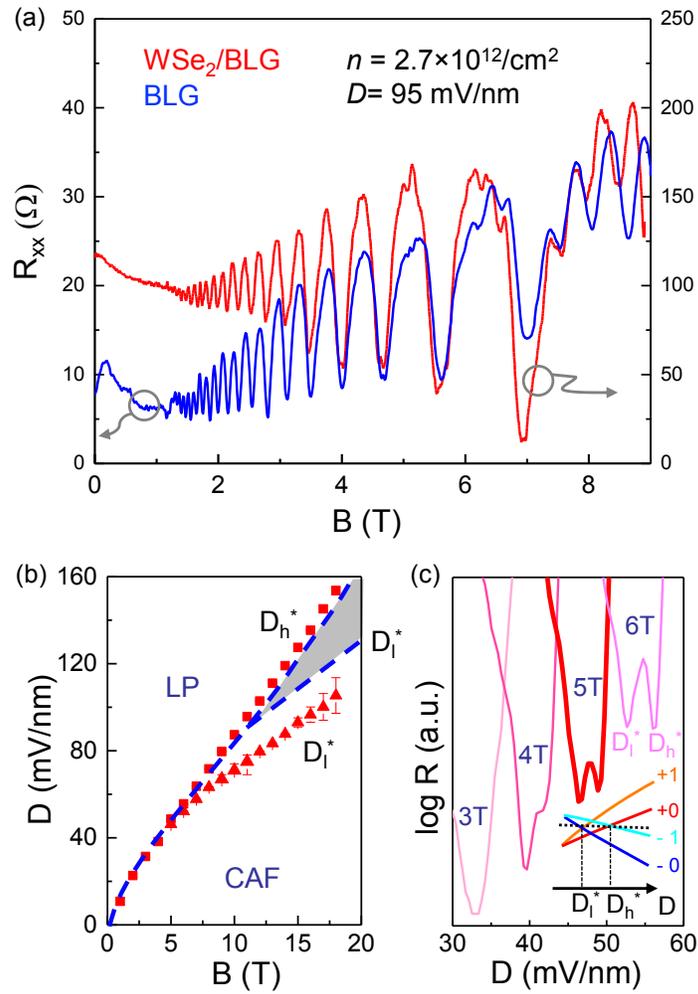

Figure 3

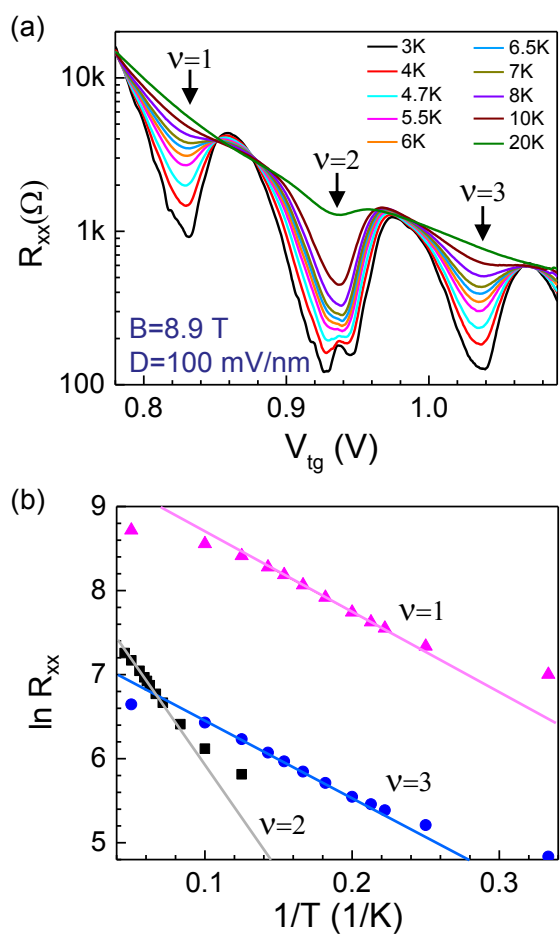

Figure 4

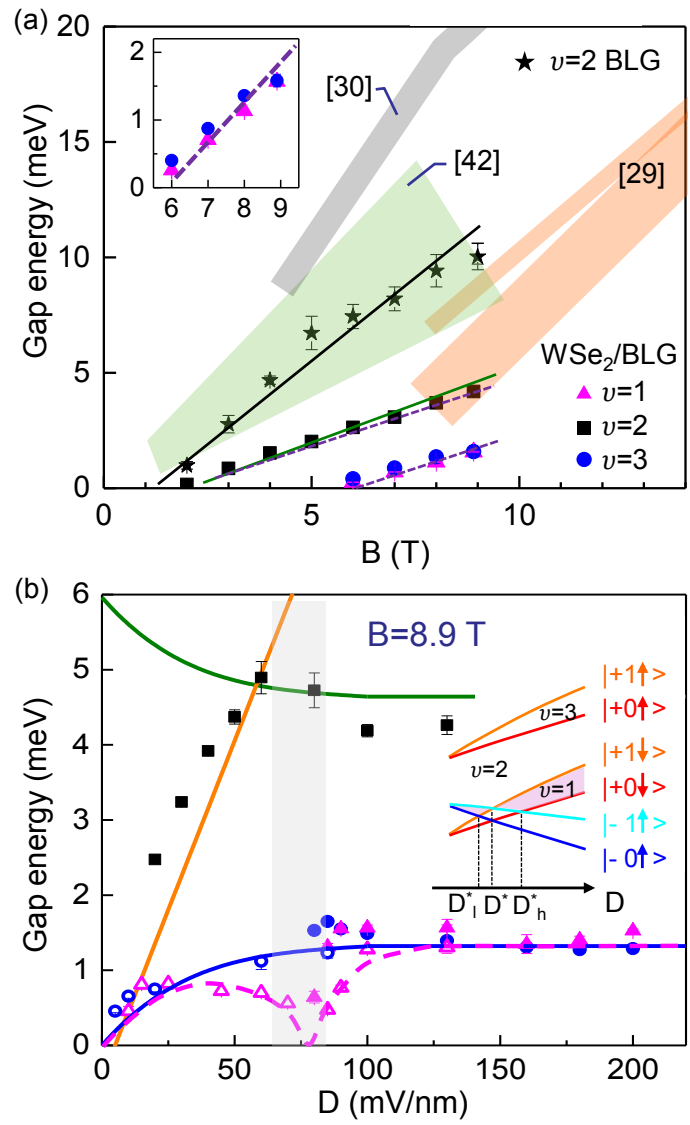

Figure 5

(a)

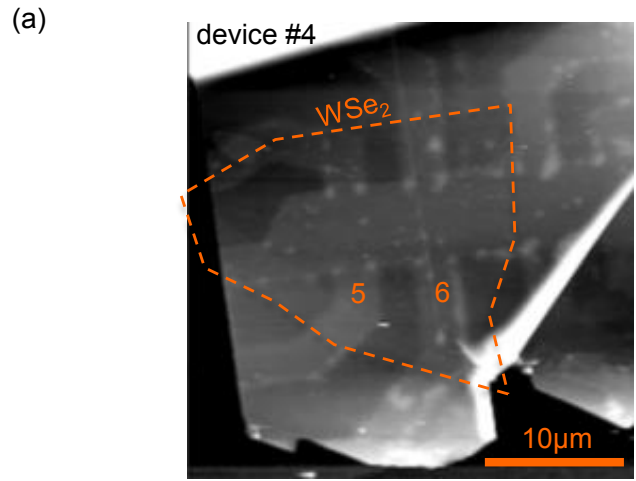

(b)

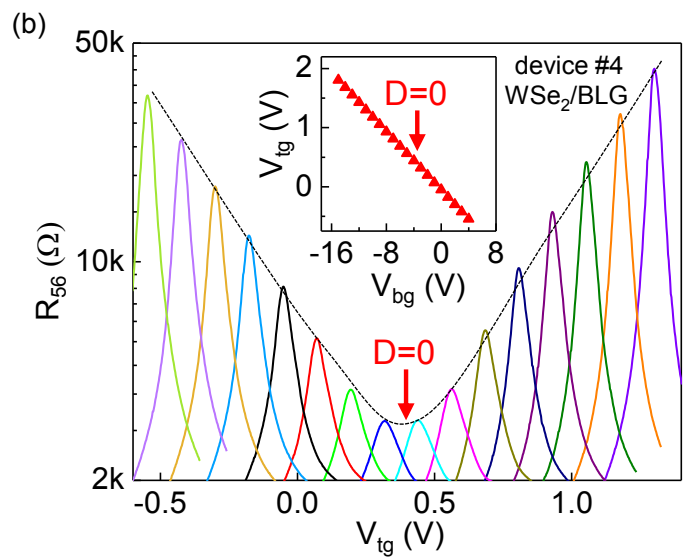

Figure 6

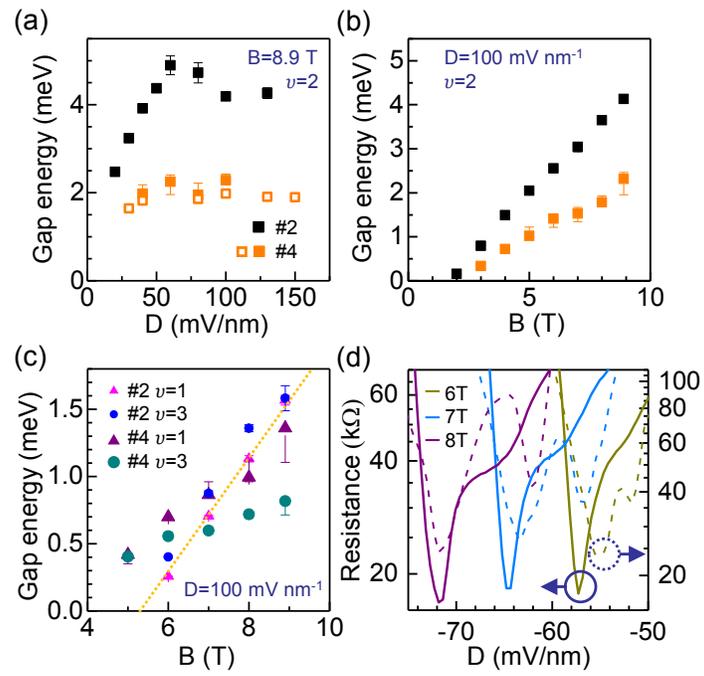

Figure 7

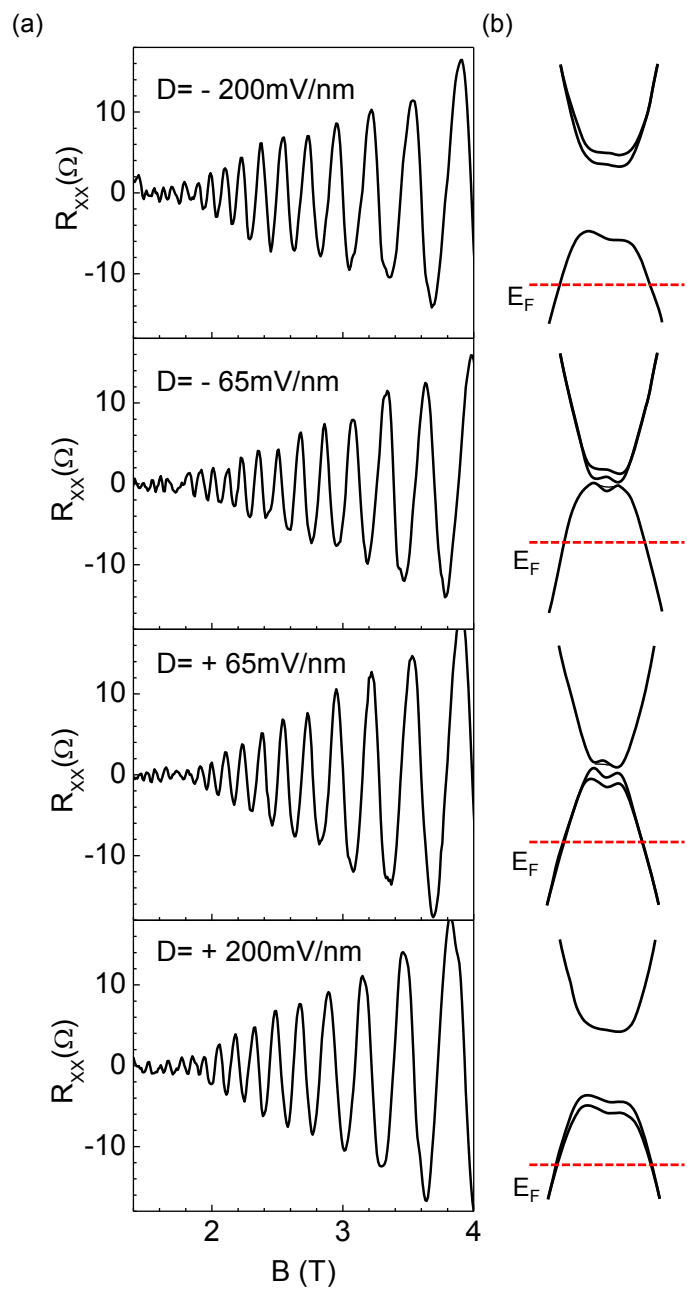

Figure 8